\documentclass[lettersize,journal]{IEEEtran}
\usepackage{amsmath,amsfonts}
\usepackage{algorithmic}
\usepackage{algorithm}
\usepackage{array}
\usepackage[caption=false,font=normalsize,labelfont=sf,textfont=sf]{subfig}
\usepackage{textcomp}
\usepackage{stfloats}
\usepackage{url}
\usepackage{verbatim}
\usepackage{graphicx}
\usepackage{cite}
\begin{document}

\title{Enviro-IoT: Calibrating Low-Cost Environmental Sensors in Urban Settings}

\author{
Thomas Johnson
Kieran Woodward
        % <-this % stops a space
\thanks{}% <-this % stops a space
\thanks{}

}

\markboth{}%
{Shell \MakeLowercase{\textit{Johnson et al.}}: Enviro-IoT: Calibrating Low-Cost Environmental Sensors in Urban Settings}

\IEEEpubid{0000--0000/00\$00.00~\copyright~2021 IEEE}

\maketitle

\begin{abstract}
Low-cost miniaturised sensors offer significant advantage to monitor the environment in real-time and accurately. The area of air quality monitoring has attracted much attention in recent years because of the increasing impacts on the environment and more personally to human health and mental wellbeing. Rapid growth in sensors and Internet of Things (IoT) technologies is paving the way for low-cost systems to transform global monitoring of air quality. Drawing on 4 years of development work, in this paper we outline the design, implementation and analysis of \textit{Enviro-IoT} as a step forward to monitoring air quality levels within urban environments by means of a low-cost sensing system. An in-the-wild study for 9-months was performed to evaluate the Enviro-IoT system against industry standard equipment is performed with accuracy for measuring Particulate Matter 2.5, 10 and Nitrogen Dioxide achieving 98\%, 97\% and 97\% respectively. The results in this case study are made up Of 57, 120 which highlight that it is possible to take advantage of low-cost sensors coupled with IoT technologies to validate the Enviro-IoT device against research-grade industrial instruments. 
\end{abstract}

\begin{IEEEkeywords}
Embedded Sensors, Air Quality, Environment, Technology, Internet-of-Things. 
\end{IEEEkeywords}

\section{Introduction}
Over the past decade, with the world becoming more compact there has been an ever significant exposure to environmental factors which are having a negative impact on health \cite{Lee2014}, \cite{WorldHealthOrganisation2022}, behaviour \cite{Mehta2012}, \cite{UrbanWellbeing23} and mental wellbeing \cite{Johnson2023Computational}, \cite{Johnson2023}. For this, air quality monitoring is becoming increasingly more common with national and international guidelines of safe levels to be met to provide a more healthier life. The equipment that is currently used to undertake this work can have hugely expensive procurement costs and factors such as installation and regular maintenance, which must be considered. For example, the purchase and installation of a particulate matter system in an existing building with power and internet connection available can cost anywhere between £10,000 and £25,000 \cite{DEFRA2006}. This cost can be increased further if maintenance visits are required for the smooth running of the site. 

The Department for Environment Food and Rural Affairs (DEFRA) currently operates the Automatic Urban and Rural Network (AURN), which is the UK's current largest air quality monitoring system that is specifically used for compliance reporting against local, national and international guidelines \cite{DepartmentforEnvironment2018}. The approach has over 274 sites and are placed in areas where air quality is poor and above the national guidelines. The collected data is highly accurate having being screened three-times prior to be made public so that any issues can be resolved \cite{DEFRAValidation}. Despite this, the system is costly and cumbersome as is based in a fixed location. Alternatively, another highly popular fixed environmental monitoring system is the AQMesh which is extremely flexible in its ability to target as little as 1 to a possible 13 different pollutants including NO, NO2, CO, SO2, H2S, CO2, PM1, PM2.5, PM4, PM10, Noise and temperature \cite{Cummins}, \cite{AQMesh2022}. The Newcastle Urban Observatory are currently using this device across the city to measure a range of environmental factors \cite{MeshNewcastle}.

The area of environmental monitoring using low-cost affordable sensors continues to receive increasing attention due to their ability to collect data continually and in real-time \cite{DeBord2016} helping to reveal early health conditions \cite{Nieuwenhuijsen2014}. Recent work in indoor air quality monitoring systems developed a custom-built single-board computer measuring a range of particulates and gases demonstrating a high correlation between industry standard equipment and specifically PM2.5 and PM10 \cite{Zhang2021}. Gualtieri et al. \cite{Gualtieri2017} further highlights the potential use of low-cost sensors by developing a road and traffic air quality management system for use in urban space which is highly accurate. It is noted that in some cases, low-cost resources often have short limited lifespans and can become less reliable over a few months \cite{Clements2017}. 

Previous work has explored the capabilities of low-cost sensors coupled with Internet of Things technologies (IoT) due to the ability of collecting data in a practical approach where it is difficult to access certain locations for access (e.g. lamp-posts) \cite{IoTSystemAir}. Additionally, using IoT in air quality monitoring helps to reduce the human-to-human or human interaction to the computer which as a result produces a continuous stream of data collection \cite{Irawan2021}. 

Over the last decade there have been few works in the literature that explore the use of embedded miniaturised computers that develop air quality sensing devices to monitor the environment outdoor in real-time, at a low-cost and validated in a real-world study. To overcome these challenges we have developed, \textit{Enviro-IoT} and present the key contributions and significance of this work as:

\begin{itemize}
    \item Design and develop a low cost sensing system for air quality monitoring called Enviro-IoT, which is equipped with an array of sensors including: Particulate Matter (PM), Nitrogen Dioxide (NO2), Oxidising Gases, Reducing Gases and Noise. 

\newpage

    \item Deployment in the real-world and calibrated against a DEFRA AURN air quality monitoring station demonstrating high accuracy of over 97\% to all pollutants. 
\end{itemize}

The key motivations of this letter aims to accelerate the deployment of low-cost environmental monitoring stations for environmental science research. It draws on a 9-month study through exploring, developing, testing and deploying a range of miniature sensors and IoT technologies within the urban environment. The rest of this letter is organised as follows: Section II presents the project overview by introducing the technology hardware used for the co-location project and discusses the device's deployment into the real-world. Section III details the results and discuses the significance. Finally, Section IV concludes this letter.

\section{Project Overview}
The project aims to propose a viable alternative solution to monitoring air quality levels through embedded low-cost sensing technologies and the use of IoT to measure impact in the urban environment in real-time. The work formed part of a 'co-location' project working alongside Nottingham City Council, Environmental Health Teams, Department for Environment, Food and Rural Affairs (DEFRA) and Ricardo Air Quality Specialists. The pollutants of particular importance for Nottingham is both the monitoring of Particulate Matter (PM) 2.5, 10 and Nitrogen Dioxide \cite{NCC2023}.

\subsection{The Design of Enviro-IoT}
\begin{figure}[h]
\centering
  \includegraphics[width= 9.3cm]{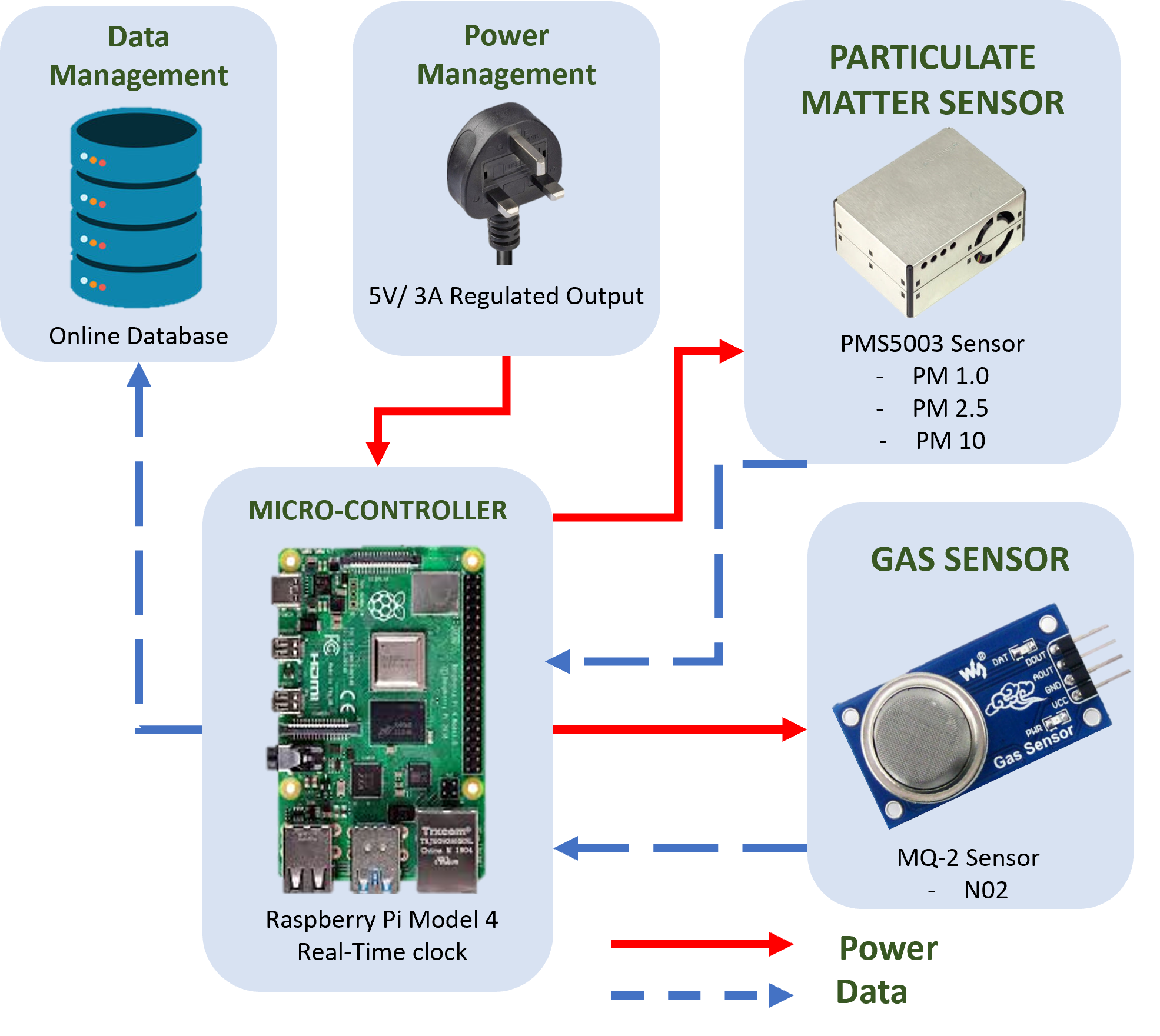}
  \caption{Module hardware interfaces of the proposed Enviro-IoT device}
  \label{fig:hardware interfaces}
\end{figure}

In this work we propose the design and implementation of the Enviro-IoT system. The proposed hardware interfaces of the low-cost air quality system is depicted at Figure \ref{fig:hardware interfaces}. This work comprises of a Raspberry Pi 4 as the main controller board. This board also utilises a sensors for the collection of Particulate Matter 1.0, 2.5 \& 10 and Nitrogen Dioxide. All components are secured within a 3D printed box measuring 12cm high, 14cm wide and 10cm depth, and is very flexible in both position and ability to monitor where space is a limited factor within the environment.

The price of each station 'Enviro-IoT' taking into consideration all components required is around £250.00. The resources can easily be acquired from UK or overseas sellers. Table \ref{components}, lists all of the key components that are required to replicate building the Enviro-IoT devices.

\begin{table}[h!]
\centering
\caption{Key components and unit costs for the Enviro-IoT. .\label{components}}
\begin{tabular}{|p{5cm}|p{2.5cm}|}
\hline \textbf{Component} &\textbf{ Cost (£ GBP)}  \\
\hline Raspberry Pi 4 board and Power Connection & £50.00-60.00   \\
\hline 3D Printed case & £15.00-25.00 \\
\hline Particulate Matter Sensor & £20.00-30.00 \\
\hline Analog gas Sensor & £18.48 - 22.00 \\
\hline Nitrogen Dioxide Sensor & £25.00 - 38.00 \\
\hline Analog to digital converter  & £10.00 - 20.00 \\
\hline
\end{tabular}
\end{table}

\subsection{Device Implementation}

\begin{figure}[h]
\centering
  \includegraphics[width= 9cm]{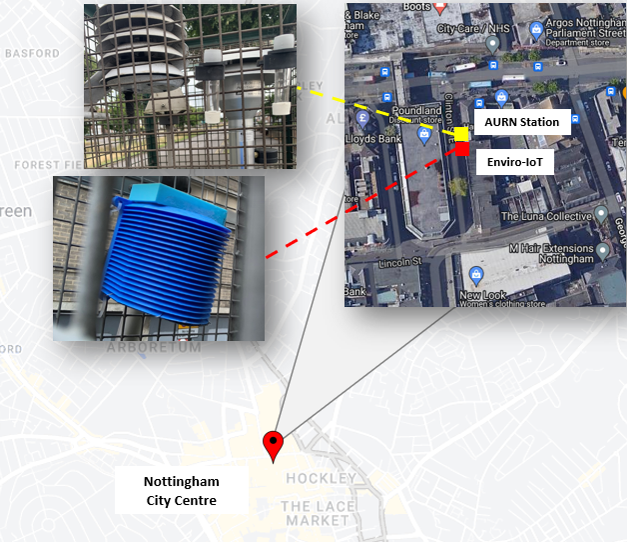}
  \caption{Enviro-IoT: Low-cost air quality monitoring device in-situ in Nottingham City Centre alongside the DEFRA Automatic Urban and Rural Network quality station.}
  \label{fig:deploymentnotts}
\end{figure}

The proposed low-cost air quality monitoring device \textit{Enviro-IoT} proposed in this work is shown at Figure \ref{fig:deploymentnotts}. Additionally, the Figure depicts the location of both the Enviro-IoT and DEFRA AURN in Nottingham City Centre which is where the device was installed for the co-location experiment. The location is situated in a busy urban area of Nottingham, U.K. 

\begin{figure*}
\includegraphics[width= 19cm, height=8cm]{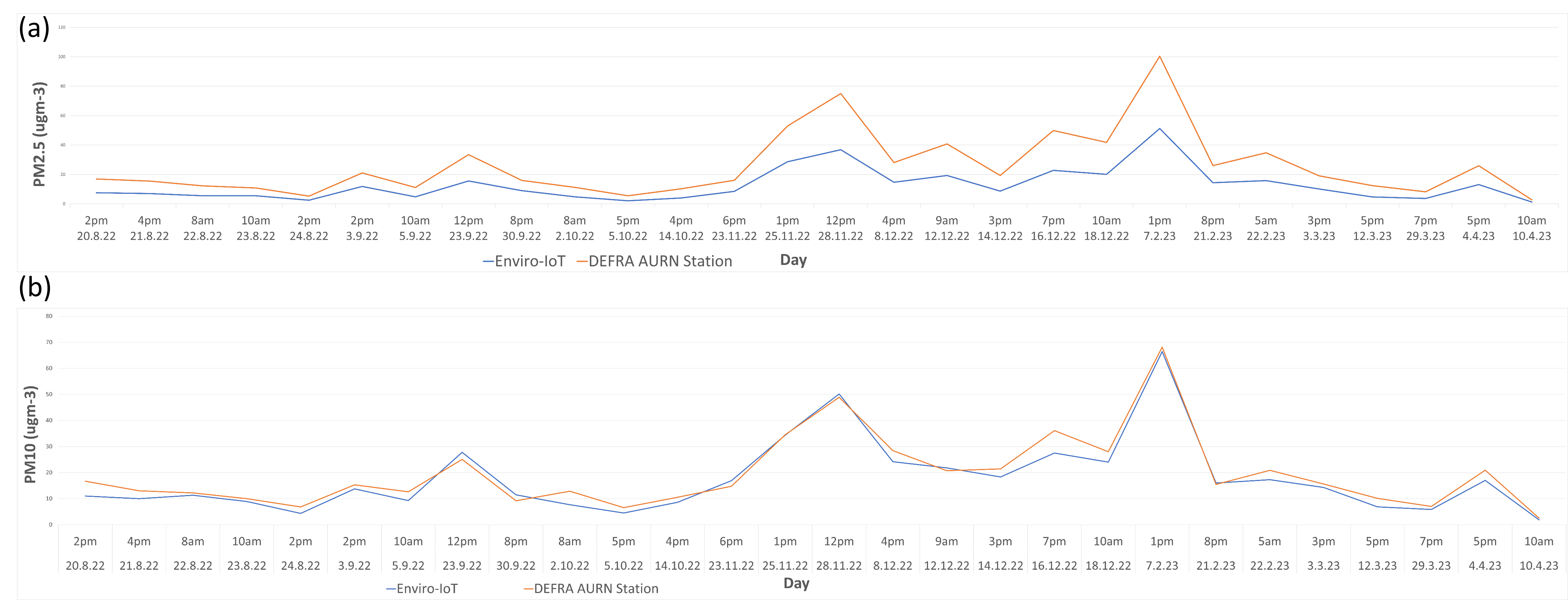}
  \caption{Performance of Enviro-IoT and DEFRA AURN for Particulate Matter 2.5 (a) and Particulate Matter 10 (b) during the 9-month study.}
  \label{fig:graphpm}
\end{figure*}

The areas around this site was selected as part of a pilot study working with Nottingham City Council and the added interest of monitoring {NO2} and {PM} levels of air pollution. The Enviro-IoT were fitted on top of the AURN with cables thread through an inlet to access power and network connections. Due to the location of sensors, the Enviro-IoT was placed into a large wire cage to protect them against vandalism. The devices have been placed close as possible to the other sensors on the AURN.

The Enviro-IoT system was installed and active between July 2022 and April 2023. This was due to the DEFRA site re-locating to a new area in Nottingham. During the installation process, there were issues around the connectivity of devices between the Enviro-IoT and online database. This was down to the free Council WiFi positioned across the City Centre only allowing devices to be connected for a maximum time of 30-minutes before being disconnected. As a resolution for this co-location project a mini-router with 4G enabled network was connected to the device to bridge the connection between the Enviro-IoT and the database. We used a 50 GB sim card within the router and this lasted for the entire 9-month study. Specifically, the collected data was PM2.5, PM10, NO2 and a timestamp. The Enviro-IoT sampled the sensor data 10 times across a 60-minute (1 hour) period and then averaged to give the mean concentration per hour to match the frequency sample reported by the AURN station, which would allow for later correlations. As a result, from a 34-week constant recording a total of 57, 120 samples were collected from the Enviro-IoT. 

\section{Results and Discussion}
\begin{figure*}
\includegraphics[width= 18cm, height=4.5cm]{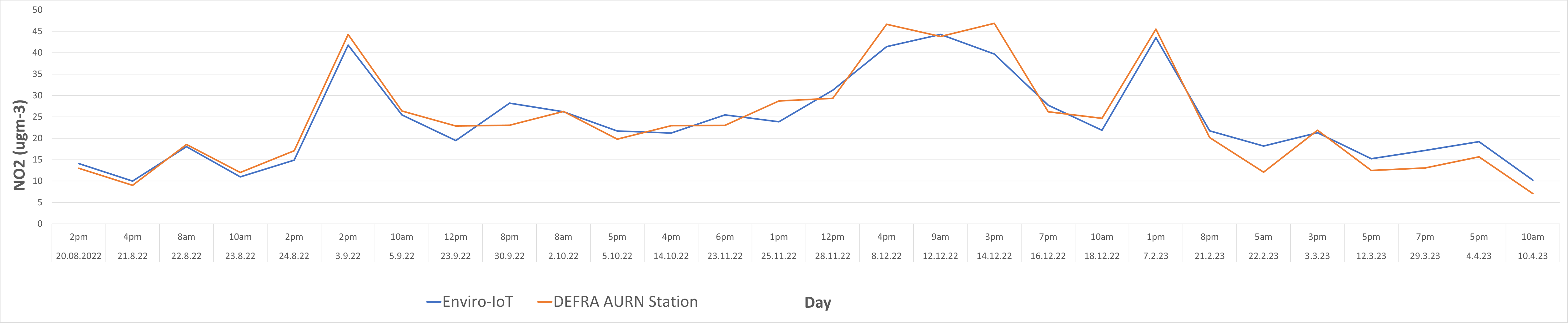}
  \caption{Performance of Enviro-IoT and DEFRA AURN for Nitrogen Dioxide during the 9-month study.}
  \label{fig:graphno}
\end{figure*}

To assess and compare the reliability of the Enviro-IoT we have collected the average hourly pollutant concentration for PM2.5, PM10 and NO2 for both devices and plotted across the 9-month study as depicted in Figures \ref{fig:graphpm} (a), (b) and \ref{fig:graphno}. As a result of data gathering issues, we have omitted the values recorded in both PM2.5, PM10 and NO2 from the January values. To assess changes across the month, we have selected a variety of times to assess changes between the two devices. 

Analysing each graph at Figure \ref{fig:graphpm} (a), (b) demonstrates that for both PM2.5 and PM10 the Enviro-IoT and DEFRA AURN with both having similar values and path lines following the same trajectory for each pollutant. In all graphs, the January values have been omitted as the Enviro-IoT encountered WiFi connection issues whereby the sim-card ran out of data so the air quality data for this month was not recorded by the system. In addition, Figure \ref{fig:graphno} presents the analysis of NO2 concentration levels recorded by both devices that similarly exhibit similar patterns over time. However, there seems to be slightly more variability and deviation from the expected trend between the two devices, especially noticeable during the months of November and December.

It is important to note that in colder months, such as November and December are generally associated with higher concentration and poor air quality levels \cite{Wine2022}. This pattern is evident in both Figures \ref{fig:graphpm} and \ref{fig:graphno} where both the Enviro-IoT and DEFRA AURN demonstrate an increase in high air quality concentrates during these months. This aligns with the understanding that certain weather conditions, such as temperature inversions and reduced dispersion, can lead to higher pollution concentrations in colder months. 

The correlations of collected PM2.5, PM10 \& NO2 demonstrate that both devices function at very similar levels throughout the 9-month study which suggests that the Enviro-IoT low-cost air quality monitoring device is capable of measuring air pollutants in urban environments and for real-world deployment. 

\begin{table}[h!]
\centering
\caption{Statistical Analysis of Environmental Pollution Variables Comparison between AURN and Enviro-IoT.\label{AURN-Enviro-IoT}}
\begin{tabular}{|c|c|c|}
\hline Pollution Type & Pearson's R & Spearman's  \\
\hline PM2.5 & 0.98 & 0.96   \\
\hline PM10 & 0.97 & 0.90   \\
\hline NO2 & 0.97 & 0.93 \\
\hline
\end{tabular}
\end{table}

To further evaluate the Enviro-IoT results we have performed a statistical analysis on the collected sensor data of PM 2.5, PM10 and NO2 from both devices using Pearson's R and Spearman's as depicted at Table \ref{AURN-Enviro-IoT}. 

The results overall indicate that the Enviro-IoT system is highly reliable and capable of producing accurate results for the three pollutants with a score of more than 90\%. The most significant result using Pearson's R correlation between the results of Enviro-IoT and DEFRA AURN is PM2.5 demonstrating 98\%. 

\section{Conclusion and Future Work}
In this letter, we have presented the design, implementation and comparison of \textit{Enviro-IoT} as a low-cost air quality monitoring device for use in the real-world. The result of over 98\% and 97\% for PM 2.5, PM 10 and NO2 respectively when comparing to industry standard monitoring equipment demonstrates the potential to revolutionise environmental monitoring through the utilisation of affordable low-cost sensors and on-board IoT. The innovative approach  enables the real-time capture of air quality levels, thereby offering valuable insights in areas where monitoring air quality is challenging. The work in this project addresses the need for accurate and up-to-date information on air quality, which can be crucial for identifying regions with poor air quality and facilitating targeted interventions to mitigate potential health risks. This advancement has the potential to significantly improve public health and enhance overall air quality management.

In the future we plan to add more environmental variables to the system which will allow for a more informed understanding of air quality in urban spaces. Additionally, a longer time period study and the opportunity of multiple locations will enable a further comparison into the Enviro-IoT's reliability. 

\bibliographystyle{IEEEtran}
\bibliography{sample-base}

\vfill

\end{document}